\documentclass[conference]{IEEEtran}
\usepackage[letterpaper, top=0.71in, left=0.625in, right=0.625in, bottom=1.03in]{geometry}

\IEEEoverridecommandlockouts

\usepackage{amsmath}
\usepackage{amssymb}
\usepackage{amsthm}
\usepackage{slashbox,pict2e}
\usepackage{xcolor}
\usepackage{float}
\usepackage{stfloats}
\usepackage{graphicx}
\usepackage{lineno}
\modulolinenumbers[5]
\usepackage{epstopdf}
\usepackage{graphicx,cite,multirow,rotating}
\usepackage{setspace}
\usepackage{pgfplots} 
\usepackage{tikz}
\usepackage{enumitem}   

\usepackage[utf8]{inputenc}
\usepackage{tabularx}
\usepackage{booktabs}

\usetikzlibrary{shapes,arrows}
\usepackage{varwidth}
\usetikzlibrary{shapes.geometric, arrows}
\usetikzlibrary{positioning,arrows}

\usepackage{amsfonts}
\usepackage{yfonts}
\usepackage{psfrag}
\usepackage{pifont}
\usepackage{amsfonts}
\usepackage{bbm}
\usepackage{dsfont}
\usepackage{color}
\usepackage{amsmath,stackrel,dsfont}
\usepackage{amssymb,stmaryrd}
\usepackage[draft,bookmarks=false]{hyperref}
\usepackage{algorithm}
\usepackage{algpseudocode}
\usepackage{pifont}
\modulolinenumbers[5]
\usepackage{amsmath}
\usepackage{epstopdf}
\usepackage{graphicx,multirow,rotating}
\usepackage{setspace}
\usepackage{pgfplots} 
\usepackage{tikz}

\usetikzlibrary{shapes,arrows}
\usepackage{varwidth}
\usetikzlibrary{shapes.geometric, arrows}
\usetikzlibrary{positioning,arrows}
\usepackage{amssymb}
\usepackage{amsfonts}
\usepackage{yfonts}
\usepackage{psfrag}
\usepackage{pifont}
\usepackage{amsfonts}
\usepackage{bbm}
\usepackage{dsfont}
\usepackage{subcaption}
\usepackage{color}
\usepackage{amsmath,stackrel,dsfont}
\usepackage{algorithm}
\usepackage{algpseudocode}
\usepackage{pifont}
\usepackage{titlesec}

\newcommand{\RN}[1]{%
	\textup{\uppercase\expandafter{\romannumeral#1}}%
}

\begin{document}
\titlespacing{\section}{3pt}{4pt plus 2pt minus 1pt}{3pt plus 2pt minus 1pt}
\titlespacing{\subsection}{3pt}{3pt plus 1pt minus 0pt}{2pt plus 1pt minus 0pt}
\titlespacing\subsubsection{3pt}{3pt plus 4pt minus 2pt}{0pt plus 2pt minus 2pt}
\setlength{\textfloatsep}{2pt plus 3pt minus 2pt}
\setlength{\abovecaptionskip}{2pt plus 3pt minus 2pt} 
\setlength{\belowcaptionskip}{2pt plus 3pt minus 2pt} 
\setlength{\belowdisplayskip}{3pt plus 3pt minus 2pt}
\setlength{\belowdisplayshortskip}{5pt plus 3pt minus 2pt}
\setlength{\abovedisplayskip}{3pt plus 3pt minus 2pt}
\setlength{\abovedisplayshortskip}{5pt plus 3pt minus 2pt}
\setlength{\skip\footins}{8pt plus 2pt minus 1pt}
\title{Energy-Efficient Federated Learning in Cooperative Communication within Factory Subnetworks}

\author{\IEEEauthorblockN{Hamid Reza Hashempour\IEEEauthorrefmark{1},  Gilberto Berardinelli\IEEEauthorrefmark{2},Shashi Raj Pandey\IEEEauthorrefmark{2}, and Hien~Quoc~Ngo\IEEEauthorrefmark{1}}

\IEEEauthorblockA{
\IEEEauthorrefmark{1} Centre for Wireless Innovation (CWI), Queen’s University Belfast, U.K. \\
\IEEEauthorrefmark{2} Department of Electronic Systems, Aalborg University, Denmark. \\
\{h.hashempoor, hien.ngo\}@qub.ac.uk, \{gb, srp\}@es.aau.dk}}

\markboth{IEEE ,~Vol.~?, No.~?,  ~}%
{Shell \MakeLowercase{\textit{et al.}}: Bare Demo of IEEEtran.cls for IEEE Journals}

\maketitle

\begin{abstract}
This paper investigates energy-efficient transmission protocols in relay-assisted federated learning (FL) setup within industrial subnetworks, considering latency and power constraints. In the subnetworks, devices collaborate to train a global model by transmitting their local models at the edge-enabled primary access (pAP) directly or via secondary access points (sAPs), which act as relays to optimize the training latency. We begin by formulating the energy efficiency problem for our proposed transmission protocol. Given its non-convex nature, we decompose it to minimize computational and transmission energy separately. First, we introduce an algorithm that categorizes devices into single-hop and two-hop groups to decrease transmission energy and then selects associated sAPs. Subsequently, we optimize the transmit power, aiming to maximize energy efficiency. To that end, we propose a Sequential Parametric Convex Approximation (SPCA) method to configure system parameters jointly. 
Numerical results demonstrate a significant reduction in outage probability and at least a twofold savings in total energy consumption, together with faster convergence, compared with single-hop transmission.
\let\thefootnote\relax\footnotetext{ 
The work of H. R. Hashempour and H. Q. Ngo was supported by a research grant from the Department for the Economy Northern Ireland under the US-Ireland R\&D Partnership Programme.}
\end{abstract}


\begin{IEEEkeywords}
Federated learning, energy efficiency, subnetworks, Industrial Internet of Things (IIoT).
\end{IEEEkeywords}

\IEEEpeerreviewmaketitle

\section{Introduction}\label{intro}

The Industrial Internet of Things (IIoT) holds significant promise for transforming industrial operations and driving future industrial advancements\cite{Nguyen}. Artificial intelligence (AI) has emerged as a cornerstone in realizing intelligent IIoT applications, often requiring centralized data collection and processing. However, in many practical scenarios, achieving centralized processing poses challenges due to the scale of modern IIoT networks and the requirements to maintain industrial data confidentiality\cite{Nguyen}. Federated Learning (FL) has emerged as a compelling privacy-preserving collaborative AI paradigm, particularly suited for intelligent IIoT networks. FL coordinates multiple IIoT devices and machines for AI training at the network edge without the need to share the actual data.\cite{Nguyen,Hu,Aouedi,Huang}. This is done as follows. The FL server periodically selects users to participate in each training round. These selected users compute training loss, update weights, and transmit local models to the server. Upon receiving these models, the server aggregates them and iterates the process until convergence. Thus, FL enables distributed AI training while preserving data privacy and scalability in IIoT environments \cite{Zhao,Zhan}.

Short-range, low-power in-X subnetworks, intended for deployment in entities such as robots, sensors, and production modules to replace wired control infrastructure in IIoT environments, are currently under investigation by both industry and academia\cite{Gilberto}. IIoT devices within these subnetworks utilize an FL-like approach for taming collective intelligence through collaborative learning while transmitting data to an edge server. However, challenges like signal blockages and fading, particularly from metallic machinery, as well as device mobility, lead to delays in training/inferencing in the subnetwork, conflicting with strict latency and reliability requirements.

Recent studies suggest that relay technology can address these challenges by improving wireless communication system coverage and reliability without additional power consumption \cite{relay1}. Nevertheless, battery-powered IoT devices used in FL have to carefully balance energy expenditures between local training demands and the need to upload model updates to a central server. 
 This adds significant resource consumption for the devices, alleviated otherwise through relays; thus, it necessitates efficient battery power management at the device. Integrating relays into the system complicates the interdependencies among various parameters, so a joint consideration of wireless resource allocation and computing capacity is needed.

In this study, we focus on energy-efficient communication in FL with relays within industrial subnetworks to address the above challenges. Our contributions are summarized below.

\begin{itemize}[label=$\bullet$]
    \item Firstly, we formulate the energy efficiency problem for our relay-assisted Time Division Multiple Access (TDMA) transmission protocol. Given the high coupling between variables, we partition the problem into minimizing computational and transmission energy.
    
    \item Next, we devise a method to schedule sensors into single-hop and two-hop transmission modes, each served by a corresponding relay based on their Channel State Information (CSI), to maximize the rate.
    
    \item We then tackle the critical challenge of minimizing total transmit energy given the timing constraints, i.e., tight cycle times for transmissions. To address the inherent non-convexity of this optimization problem, we employ a promising technique called Sequential Parametric Convex Approximation (SPCA).
\end{itemize}

To the best of our knowledge, previous literature has not delved into integrating a relay-assisted TDMA transmission protocol with stringent time constraints FL training within factory subnetworks.

The remainder of this paper is structured as follows: Section~\ref{Sys_Model} introduces the system model and the proposed communication protocol, while Section~\ref{Optim} presents the proposed method for energy minimization. In Section~\ref{Simulat}, simulation results are presented to validate the effectiveness of our approach. Finally, conclusions are summarized in Section~\ref{conc}.

\section{System Model}\label{Sys_Model}
In an industrial subnetwork, $N$ sensors are connect wirelessly to $K+1$ access points (APs). One AP serves as the primary AP (pAP) and controls actuators, while the remaining $K$ APs are secondary APs (sAPs) with only radio communication capabilities.
The sAPs form set $\mathcal{K}$, and the sensors are in set $\mathcal{N}$. Sensors can communicate directly with the pAP (single-hop) or both pAP and sAP (two-hop cooperative). Devices scheduled for single-hop transmission are in set $\mathcal{N}_{1h}$, and those for two-hop transmission are in set $\mathcal{N}_{2h}$.
APs and devices collaborate on an FL algorithm over wireless networks for data analysis and inference. The FL model trained by each device's dataset is termed the local FL model, while the model generated by the pAP using inputs from all devices is the global FL model.

\subsection{FL Model}
In FL, users train a statistical model locally on the dataset stored on their mobile devices. To enhance model performance, the aim of device $n$ is to minimize the training loss by optimizing the weight parameter $\mathbf{w}_n$ concerning its local dataset $\mathcal{D}_n$ as follows:
\begin{align}\label{FL-model}
& \displaystyle  \min_{\mathbf{w}_n \in \mathbb{R}} F_n(\mathbf{w}_n) = \sum_{j \in \mathcal{D}_n} l(\mathbf{w}_n,x_j,y_j).
\end{align}
In this study, the cross-entropy loss function is employed to train the learning model, which is implemented using a feedforward neural network (FNN). After local training at device $n$, the weight vector $\mathbf{w}_n$ is transmitted to the pAP for model aggregation. A global model is then constructed as defined in \cite{FL1}, characterized by
\begin{align}
  \mathbf{\Bar{w}} &  = \dfrac{\sum_{n \in \mathcal{N}}D_n\mathbf{w}_n}{\sum_{n \in \mathcal{N}}D_n},
\end{align}
where $D_n$ is the size of local data samples.
Since the data is distributed, solving \eqref{FL-model} directly is generally challenging. Therefore, FL typically employs an iterative algorithm to train a global model from the devices. Specifically, in each round, the $n$th device computes the training loss and then updates the weights using gradient descent as
\begin{align}
  \mathbf{v}_n &  =\mathbf{w}_n -\mu \nabla  F_n(\mathbf{w}_n),
\end{align}
where $\mathbf{v}_n$ represents the updated model parameter of the $n$th device, and $\mu$ denotes the learning rate. Following this, the updated local models from multiple devices are collected and aggregated at the pAP.

\subsection{Wireless Communication Model}

\begin{figure}
	\centering
		\includegraphics[width=.9\linewidth]{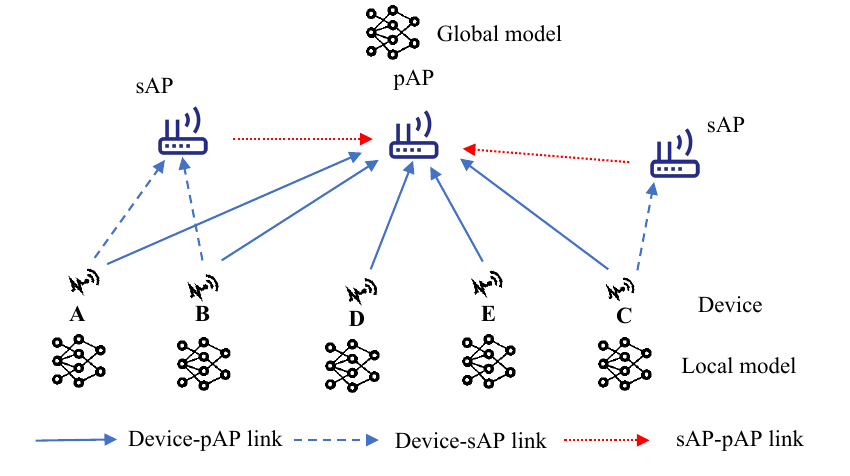}
	\caption{System model for FL with multiple APs within a subnetwork.}
	\label{fig1a}
 \end{figure}

Fig.~\ref{fig1a} shows an example subnetwork with 5 sensors and 3 APs. Sensors A, B, and C operate in the two-hop cooperative mode, while sensors D and E use the single-hop mode. In cooperative mode, the pAP receives packets from both the cooperative link and a direct link, enabling it to effectively combine the energy received from sensors and sAPs.

All APs are assumed to be time-synchronized and use a shared frequency band. Sensors are allocated time resources in a TDMA manner to prevent intra-cell interference. This study concentrates solely on the uplink (UL), where each sensor ($n=1, \cdots, N$) transmits a packet of $s$ bits to the pAP/sAPs over a bandwidth of $W$ Hz. The pAP has to receive all sensor packets within a time slot of $T$ seconds.

We assume the pAP knows all channel responses between sensors and the pAP/sAP and between sAPs and the pAP. This information is obtained during a training phase, during which channel responses are estimated at the pAP and sAPs through transmission of reference sequences, with the latter reported to the pAP. Channel responses are expected to remain stable in static sensor scenarios. Thus, in practice, the training operation can be periodically repeated at a relaxed frequency.
The pAP scheduler utilizes complete CSI from all subnetwork links to determine the subset of devices for single-hop and two-hop transmission, their transmission rates, time resource allocation, and transmit power.
We posit that communication reliability in a subnetwork can be improved by the presence of sAPs and their forwarding capabilities, necessitating a tailored communication protocol.

\begin{figure} 
	\centering
	\includegraphics[width=1\linewidth]{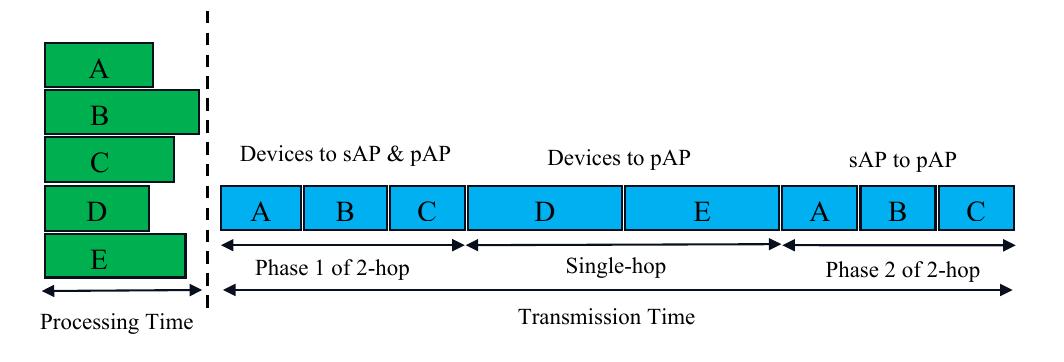}
 \caption{An implementation of an FL algorithm via proposed cooperative TDMA protocol}
	\hfill
	\label{time}
\end{figure}

 Our proposed protocol is illustrated in Fig.~\ref{time}. The UL time slot is divided into three sub-slots of variable duration, corresponding to the first phase of the two-hop transmissions, the single-hop transmissions, and the second phase of the two-hop transmission. Sensors belonging to the $\mathcal{N}_{2h}$ set then use the first phase sub-slot to transmit their packets, which is received by both the pAP and sAP. We assume that each sensor in $\mathcal{N}_{2h}$ is served  only by the sAP for which it experiences the most advantageous channel conditions. The sAP then uses the second phase sub-slot to forward the received message from the sensors to the pAP. We assume that the sAP acts as a Decode-and-Forward (DF) relay. As mentioned above, the pAP can combine the energy from the reception of the signal transmitted by sensors in $\mathcal{N}_{2h}$ during the first phase transmission with the energy received from the sAPs before decoding. Transmissions by sensors scheduled in the single-hop sub-slot are received  only by the pAP.

Below, we outline the signal model for single-hop and two-hop DF relaying.
For notation, $\mathbf{h}_c \in \mathbb{C}^{K \times 1}$ represents the vector of channel responses between the pAP and each of $K$ sAPs. The channel from the pAP to $N$ devices is denoted as $\mathbf{h}_{p} \in \mathbb{C}^{N \times 1}$, while the matrix $\mathbf{H}_{s} \in \mathbb{C}^{N \times K}$ represents the channels from $N$ devices to $K$ sAPs.

\subsubsection{Signal model for single-hop transmission}
The Signal-to-Noise Ratio (SNR) of the $n$th device in the direct link to the pAP is expressed as
\begin{align} 
g_{n}^d&=\dfrac{P_n|\mathbf{h}_p(n)|^2}{\sigma_0^2},\ \ \forall  n \in \mathcal{N}_{1h} .
\end{align}
$\sigma_0$ represents the standard deviation of the noise, and $P_n$ denotes the transmit power of device $n$. The attainable information rate of the $n$th device in the direct link to the pAP is expressed as
\begin{align}
r_{n}^d &= \mathrm{log}_2 \left( 1+ g_{n}^d  \right), \ \ \forall  n \in \mathcal{N}_{1h} .
\end{align}

\subsubsection{Signal Model for DF relaying}

We examine the cooperative link, where transmissions from devices are first received by the sAPs, decoded, and then forwarded to the pAP.
We propose a relaying scheme where each device is assigned the strongest sAP among all available ones. This strategy, termed the `$1$ of $K$' relaying method, provides a flexible framework for improving communication reliability and extending coverage.
We define the index of the strongest sAP for the $n$th device as $k^*_n$, where $k^*_n \in \mathcal{K}$. The SNR and achievable rate of the $n$th device at the $k^*_n$th sAP are given, respectively, by
\begin{align}
g_{n,k^*_n}&=\dfrac{P_n|\mathbf{H}_s(n,k^*_n)|^2}{\sigma_0^2},\ \ \forall  n \in \mathcal{N}_{2h}, \ \forall  k^*_n \in \mathcal{K},  \\
r_{n,k^*_n}^{(1)} &= \mathrm{log}_2 \left( 1+ g_{n,k^*_n}  \right), \ \ \forall  n \in \mathcal{N}_{2h},\ \forall  k^*_n \in \mathcal{K},
\end{align}
where the superscript ${(1)}$ indicates the first phase in the two-hop cooperative transmission.
Assuming that the sAP re-encodes and transmits the received signal to pAP, the SNR and the achievable rate from the $k^*_n$th sAP to the pAP  are respectively given by
\begin{align}\label{eq7}
g_{k^*_n,p}&=\dfrac{P_{k^*_n}^s|\mathbf{h}_c(k^*_n)|^2}{\sigma_0^2},\ \forall  n \in \mathcal{N}_{2h}, \ \forall  k^*_n \in \mathcal{K} ,\\
r^{(2)}_{n} &= \mathrm{log}_2 \left( 1+ g_{k^*_n,p} + g_{n}^d \right),\ \forall  n \in \mathcal{N}_{2h}, \ \forall  k^*_n \in \mathcal{K} ,
\label{eq7}
\end{align}
where the superscript ${(2)}$ denotes the second phase in the two-hop cooperative transmission. Also, $P_{k^*_n}^s$ is the transmit power of the $k^*_n$th sAP. Equation \eqref{eq7} accounts for energy combining of the two transmission phases at the pAP receiver.

\subsection{Processing and Transmission Model}
The FL process between devices and their serving pAP is depicted in Fig. \ref{fig1a}. This procedure comprises three steps at each iteration: local computation at each device (with multiple local iterations), transmission of the local FL model for each device, and aggregation and broadcast of results at the pAP. Each device computes its local FL model using its local dataset and the received global FL model during the local computation step.

\subsubsection{Local Computation}
Let $f_n$ denote the computation capacity of device $n$, quantified by the number of CPU cycles per second. The computation time required at device $n$ for data processing is given by
\begin{align}\label{tau}
    \tau_n = \dfrac{I_n C_n D_n}{f_n},\ \forall n \in \mathcal{N}.
\end{align}
Here, $C_n$ (cycles/sample) represents the number of CPU cycles needed to compute one sample of data at device $n$, and $I_n$ stands for the number of local iterations at device $n$. As per \cite{Energy1}, the energy consumption for performing a total of $C_n D_n$ CPU cycles at device $n$ is
\begin{align}
    E_{n1}^L = \kappa  C_n D_n f_n^2,
\end{align}
where $\kappa$ denotes the effective switched capacitance dependent on the chip architecture. To compute the local FL model, device $n$ must perform $ C_n D_n$ CPU cycles across $I_n$ local iterations, resulting in the total local computation energy at device $n$ as
\begin{align}
    E_{n}^L = I_n E_{n1}^L = \kappa I_n C_n D_n f_n^2.
\end{align}

\subsubsection{Wireless Transmission}
In this phase, device $n$ must transmit the local FL model to the pAP. Since the dimensions of the local FL model remain constant for all devices, each device's data size for upload remains consistent, denoted as $s$.
Considering time-division multiplexing over a bandwidth $W$, a packet of $s$ bits for the $n$th device can be transmitted within $\dfrac{s}{W r_{n}^d }$ time. Transmitting packets from all single-hop devices in a TDMA manner results in a total transmission time
\begin{align}
 T_{1h} &= \sum_{n \in \mathcal{N}_{1h}} \dfrac{s}{W r_{n}^d }.
\end{align}
To execute DF, the signal from device $n$ must be accurately decoded by the strongest sAP and then re-encoded into a new message. The duration of over-the-air time required for successful transmission of a packet containing $s$ bits by device $n$ is therefore
$\dfrac{s}{W   r_{n,k^*_n}^{(1)} }+\dfrac{s}{W  r_n^{(2)}},\ \forall  n \in \mathcal{N}_{2h}$.
Let's denote the total transmission time for all devices in the first and second phases of the two-hop method as 
\begin{align}\label{eq10}
T_{2h}^{(1)} & = \sum_{n \in \mathcal{N}_{2h}} \dfrac{s}{W  r_{n,k^*_n}^{(1)} }, \\
T_{2h}^{(2)} & = \sum_{n \in \mathcal{N}_{2h}} \dfrac{s}{W  r_n^{(2)}}.
\end{align}
The total transmission time in both single-hop and two-hop cases is then calculated as
\begin{align}\label{eq10}
T^{\mathrm{UL}}&=T_{1h} +T_{2h}^{(1)} + T_{2h}^{(2)}, 
\end{align}
To transmit data of size $s$ within a time duration $T^{UL}$, the wireless transmit energy will be given by
\begin{align}\label{eq10}
E^T & =\dfrac{s}{W} \left [\sum_{n \in \mathcal{N}_{1h}} \dfrac{P_n}{ r_{n}^d }+\sum_{n \in \mathcal{N}_{2h}}\left (\dfrac{P_n }{   r_{n,k^*_n}^{(1)} }
+\dfrac{P_{k^*_n}^s }{r_n^{(2)}}\right)\right].
\end{align}

Considering the FL model outlined above, each user's energy consumption comprises both local computation energy $E_{n}^L$ and wireless transmission energy $E^T$. Let's denote the number of global iterations as $I_0$. Then, the total energy consumption of all users participating in FL will be
\begin{align}
E & = I_0 \left( E^T + \sum_{n \in \mathcal{N}} E_{n}^L  \right)
\end{align}

Each device $n \in \mathcal{N}$ possesses a local dataset. Hereafter, the total time $T^c$ required to execute the FL algorithm is referred to as the completion time. The completion time of each user includes both local computation time and transmission time, as illustrated in Fig. \ref{time}. Based on \eqref{tau}, the completion time $T_n^c$ of device $n$ will be
\begin{align}
T_n^c & = I_0  \left(  \tau_n  +T^{\mathrm{UL}} \right)
\end{align}
Let $T^{\mathrm{th}}$ be the maximum allowable  time for training the entire
FL algorithm, then we have $T_n^c \leq T^{\mathrm{th}} $.
Thus, the system outage probability of the FL is given
by
\begin{align}
P_{out} = \dfrac{1}{N}\sum_{n \in \mathcal{N}} \mathrm{Pr} [T_n^c > T^{\mathrm{th}}].
\end{align}

\section{Proposed Method for Energy Minimization}\label{Optim}
In this section, we establish the energy minimization problem for FL. Given the nonconvex nature of the problem, obtaining the globally optimal solution is challenging. Therefore, we propose a low-complex iterative algorithm to address the energy minimization problem.
\subsection{Problem Formulation} 
Our objective is to minimize the total energy consumption of all users while adhering to a latency constraint. This energy-efficient optimization problem can be formulated as follows:
\begin{subequations} \label{P0}
	\allowdisplaybreaks
	\begin{align} &\min_{\mathbf{P},\mathbf{f}, \mathcal{N}_{1h},\mathcal{N}_{2h},\mathcal{D}} E, \label{P0-a}\\
	\mathrm{s.t.}\ & 
I_0 \left(    \dfrac{I_n C_n D_n}{f_n}   +T^{\mathrm{UL}} \right)\leq  T^{\mathrm{th}}, \label{P0-b}
\\& 0 \leq P_n \leq P_{\max}, \ \forall n \in \mathcal{N}\label{P0-c}\\
	& 
	0 \leq P_{k^*_n}^s \leq P_{\max}, \ \forall n \in \mathcal{N}_{2h}\label{P0-d}\\
 & 
	 0 \leq f_{n} \leq f_{n}^{\max}, \ \forall n \in \mathcal{N}\label{P0-e}	
	\end{align} 
\end{subequations}
where $\mathbf{P} = [P_1, P_2, \ldots, P_N, P_{k^*_1}^s, P_{k^*_2}^s, \ldots, P_{k^*_{N_{2h}}}^s]$ represents the vector of transmit powers for all devices and sAPs cooperating in transmission. $P_{\text{max}}$ and $f_{n}^{\max}$ represent the maximum allowable transmission power and the maximum computational capacity of device $n$, respectively. Constraint \eqref{P0-b} addresses the requirement for low latency, while \eqref{P0-c} and \eqref{P0-d} set the power limits.

The formulated energy efficiency problem poses a challenge due to its non-convexity and the strong coupling among decision variables. Therefore, we initially decouple the problem into the computing resource management problem and the transmit energy efficiency problem.

\subsection{Designing the Device CPU Frequency}\label{CPU Freq}
Initially, we optimize the frequency once the number of training iterations required to achieve a specific accuracy is known. Minimizing the total energy consumption across all devices is equivalent to minimizing the individual energy consumption of each device. For each device, $E_{n}^L$ is an increasing function with respect to $f_n$. The time constraint \eqref{P0-b} of Problem \eqref{P0} implies that each device should operate at the lowest frequency $f_n^*$ permitted by the delay constraint. Thus, we obtain
\begin{align}
    f_n^* = \dfrac{I_n C_n D_n}{T^{\mathrm{th}} /I_0 -T^{\mathrm{UL}} }. 
\end{align}

\subsection{Proposed Method for Relay Selection and Transmit Power Control}\label{Optim_relay_power}
After minimizing the computation energy $E_{n}^L$, the formulated problem transforms into the energy transmission efficiency problem, expressed as follows:
\begin{subequations} \label{P1}
	\allowdisplaybreaks
	\begin{align} &\min_{\mathbf{P}, \mathcal{N}_{1h},\mathcal{N}_{2h},\mathcal{D}} \left [\sum_{n \in \mathcal{N}_{1h}} \dfrac{P_n}{ r_{n}^d }+\sum_{n \in \mathcal{N}_{2h}}\left (\dfrac{P_n }{   r_{n,k^*_n}^{(1)} }
+\dfrac{P_{k^*_n}^s }{r_n^{(2)}}\right)\right], \label{P1-a}\\
	\mathrm{s.t.}\ & 
  T^{\mathrm{UL}}\leq  T', \label{P1-b}
\\& 0 \leq P_n \leq P_{\max}, \ \forall n \in \mathcal{N}\label{P1-c}\\
	& 
	0 \leq P_{k^*_n}^s \leq P_{\max}, \ \forall n \in \mathcal{N}_{2h}\label{P1-d}
	\end{align} 
\end{subequations}
where $T'$ is the transmission time requirement  obtained from \eqref{P0-b}. Considering the complexity of jointly selecting relays and minimizing   transmission energy, our approach initially identifies the optimal transmission link before minimizing the transmit energy. Assuming constant transmit power, the objective is to select the link that maximizes the achievable rate, thereby minimizing delay. Under equal power allocation, i.e., $P_n = P_{k_n^*}^s$, the performance of a two-hop DF link is limited by its weakest hop. 
Thus, the effective channel gain is defined as the minimum gain over the two transmission phases. For all $n \in \mathcal{N}$ and $k \in \mathcal{K}$, we define \cite{Hashempour}
\begin{align}
\mathbf{G}^{2h}(n,k) &\triangleq \frac{1}{2} \min \left\{|\mathbf{h}_c(k)|^2, |\mathbf{H}_s(n,k)|^2 \right\}, \\
\mathbf{g}^{2h}(n) &\triangleq \max_k \mathbf{G}^{2h}(n,k).
\label{16}
\end{align}
This effective gain is compared with the direct channel gain $|\mathbf{h}_p(n)|^2$, which reflects the achievable rate under the same power constraint. If the two-hop gain exceeds the direct gain, cooperative transmission is selected; otherwise, direct transmission is preferred. In the cooperative case, the sAP corresponding to the maximum effective gain is chosen. The detailed procedure is outlined in Algorithm~\ref{Alg1}.
\begin{algorithm}
	\caption{Algorithm for classification of devices and relay selection}
	\label{Alg1}
	\begin{algorithmic}[1]
		\State \textbf{Input:}
		The channel gains $\mathbf{h}_c(k)$, $\mathbf{H}_s(n,k )$, and $\mathbf{h}_p(n)$    for all $N$ devices and $K$ sAPs.
		\For {$n= 1: N$} 
		\State calculate $\mathbf{g}^{2h}(n)$ from  \eqref{16}
            \If{$|\mathbf{h}_p(n)|^2 \geq \mathbf{g}^{2h}(n)$}
                \State $n \rightarrow \mathcal{N}_{1h}$ 
             \Else
                 \State $n \rightarrow \mathcal{N}_{2h}$
                 \State $ \arg \max_k \{\mathbf{G}^{2h}(n,:)\} \rightarrow k^*_n$
            \EndIf
		\EndFor
		\State Output: $\mathcal{N}_{1h}$, $\mathcal{N}_{2h}$, $\mathcal{K}^*$.
	\end{algorithmic}
\end{algorithm}

Given the sets $\mathcal{N}_{1h}$, $\mathcal{N}_{2h}$, and $\mathcal{D}$, we proceed to minimize power consumption. This is achieved by simplifying the optimization problem outlined in \eqref{P1} as follows:
\begin{subequations} \label{P2}
	\allowdisplaybreaks
	\begin{align} &\min_{\mathbf{P}} \left [\sum_{n \in \mathcal{N}_{1h}} \dfrac{P_n}{ r_{n}^d }+\sum_{n \in \mathcal{N}_{2h}}\left (\dfrac{P_n }{   r_{n,k^*_n}^{(1)} }
+\dfrac{P_{k^*_n}^s }{r_n^{(2)}}\right)\right], \label{P2-a}\\
	\mathrm{s.t.}\ & 
\sum_{n \in \mathcal{N}_{1h}} \dfrac{s}{W	 \mathrm{log}_2 \left( 1+ {g}_{n}^d  \right)} \label{P2-b}\\ \nonumber & + \sum_{n \in \mathcal{N}_{2h}} \dfrac{s}{W  \displaystyle \mathrm{log}_2 \left( 1+ {g}_{n,k^*_n}  \right) }\\ \nonumber& +\sum_{n \in \mathcal{N}_{2h}} 
\dfrac{s}{W  \mathrm{log}_2 \left( 1+  {g}_{k^*_n,p}  + {g}_n^d\right)}\leq  T', 
\\&
\eqref{P1-c}-\eqref{P1-d},	
	\end{align} 
\end{subequations}
where \eqref{P2-b} refers to the total time constraint. 

Problem \eqref{P2} is hard to solve because of the non-convex constraint \eqref{P2-b}, and thus, finding the global optimum is typically intractable due to the non-convex nature of the problem. To address this, we employ Sequential Parametric Convex Approximation (SPCA), where a sequence of convex programs iteratively approximates the problem. At each iteration, the non-convex constraint is replaced by a convex surrogate, serving as an approximation.
Consequently, we can express \eqref{P2} as follows:
\begin{subequations} \label{P3}
	\allowdisplaybreaks
	\begin{align} 	&\min_{\mathbf{P},\mathbf{\omega},\mathbf{t}} \ 
 E_t
 \label{P3-a}\\&
 \mathrm{s.t.}\
 \left [\sum_{n \in \mathcal{N}_{1h}}  \dfrac{1}{t_n^{(1)}} +\sum_{n \in \mathcal{N}_{2h}}\left (\dfrac{1}{t_n^{(2)}} +\dfrac{1}{t_n^{(3)}} \right)\right] \leq E_t,  \label{P3-a1}\\ & 
       \dfrac{(\omega_n^d)^2}{P_n} \geq t_n^{(1)}, \ \forall  n \in \mathcal{N}_{1h}
         \label{P3-b}\\&
          \gamma_{n}^d \geq (\omega_n^d)^2, \ \forall  n \in \mathcal{N}_{1h}
          \label{P3-c}\\&
          \dfrac{(\omega_n^{(1)})^2}{P_n} \geq t_n^{(2)}, \ \forall  n \in \mathcal{N}_{2h}
         \label{P3-d}\\&
          \gamma_{n,k^*_n}^{(1)} \geq (\omega_n^{(1)})^2, \ \forall  n \in \mathcal{N}_{2h}
          \label{P3-e}\\&
          \dfrac{(\omega_n^{(2)})^2}{P_{k^*_n}^s} \geq t_n^{(3)}, \ \forall  n \in \mathcal{N}_{2h}
         \label{P3-f}\\&
          \gamma_{n}^{(2)} \geq (\omega_n^{(2)})^2, \ \forall  n \in \mathcal{N}_{2h}
          \label{P3-g}\\&
      \sum_{n \in \mathcal{N}_{1h}} \dfrac{1}{\gamma_n^d} +\sum_{n \in \mathcal{N}_{2h}}  \left ( \dfrac{1}{\gamma_{n,k^*_n}^{(1)}}+\dfrac{1}{ \gamma_{n}^{(2)}} \right)
\leq  \dfrac{T' W}{s},\label{P3-h} 
\\&
\mathrm{log}_2 \left( 1+ g_{n}^d  \right) \geq \gamma_{n}^d, \ \ \forall  n \in \mathcal{N}_{1h},\label{P3-i}
\\&
\mathrm{log}_2 \left( 1+ g_{n,k^*_n}  \right) \geq \gamma_{n,k^*_n}^{(1)}, \  \forall  n \in \mathcal{N}_{2h},   \label{P3-j}
\\&
\mathrm{log}_2 \left( 1+  g_{k^*_n,p}  + {g}_{n}^d \right) \geq \gamma_{n}^{(2)}, \   \forall  n \in \mathcal{N}_{2h},\label{P3-k}
\\&
\eqref{P1-c}-\eqref{P1-d}	
	\end{align} 
\end{subequations}
where  $E_t$ is the energy efficiency metric, $t_n^{(1)}$, $t_n^{(2)}$, $t_n^{(3)}$, $\omega_n^{d}$, $\omega_n^{(1)}$, $\omega_n^{(2)}$, $\gamma_n^d$, $\gamma_{n,k^*_n}^{(1)}$, and $\gamma_{n}^{(2)}$ are auxiliary variables to approximate the non-convex terms with convex counterparts. 
It can be perceived that $\gamma_n^d$, $\gamma_{n,k^*_n}^{(1)}$, and $\gamma_{n}^{(2)}$ play the roles of lower bound for $\mathrm{log}_2 \left( 1+ {g}_{n}^d  \right)$, $\mathrm{log}_2 \left( 1+ {g}_{n,k^*_n}  \right)$, and $\mathrm{log}_2 \left( 1+ {g}_{k^*_n,p}  + {g}_{n}^d  \right)$, respectively. 
Increasing the lower-bound values and simultaneously
reducing the upper bounds will boost the left side of
the constraints, which are needed here, so that the constraints
\eqref{P3-h}-\eqref{P3-k} would be active at the optimum.
The \eqref{P3-h} is convex since it is a linear combination of three quadratic terms over linear functions that is convex \cite{CVX}. 
The left side of \eqref{P3-b}, \eqref{P3-d}, and \eqref{P3-f} are noncovex. To get rid of nonconvexity, we define the function $\Omega^{[i]}(\omega,z) $, as the
first-order lower approximation of them as follows:
\begin{align}\label{omega_approx}
  \dfrac{\omega^2}{z}\geq \dfrac{2\omega^{[i]}}{z^{[i]}}\omega - (\dfrac{\omega^{[i]}}{z^{[i]}})^2 z \triangleq
  \Omega^{[i]}(\omega,z),
\end{align}
where $(\omega^{[i]},z^{[i]})$ are the values of the variables $(\omega, z)$ at the
output of the $i$th iteration.
Affine approximations of constraints \eqref{P3-i}-\eqref{P3-k},  are given by
\begin{subequations}\label{lem1-formul}
	\allowdisplaybreaks
         \begin{align}
	&1 + \rho_{n} - 2^ {\gamma_n^d}  \geq 0,\ \forall  n \in \mathcal{N}_{1h},
	\\&
 \rho_{n} \leq \dfrac{P_n|\mathbf{h}_p(n)|^2}{\sigma_0^2},\ \forall  n \in \mathcal{N}_{1h},
	\\&
	1 + \psi_{n} - 2^ {\gamma_{n,k^*_n}^{(1)} }  \geq 0,\ \forall  n \in \mathcal{N}_{2h}, 
	\\&
	 \psi_{n} \leq \dfrac{P_{n}
\displaystyle |\mathbf{H}_s(n,k^*_n) |^2}{\sigma_0^2},\ \forall  n \in \mathcal{N}_{2h} \\&
1 + \zeta_{n} - 2^ {\gamma_{n}^{(2)} }  \geq 0,\ \forall  n \in \mathcal{N}_{2h}, 
	\\&
	\zeta_{n} \leq  \dfrac{P_{k^*_n}^s|  \mathbf{h}_c(k^*_n)|^2
 +P_{n}|\mathbf{h}_p(n)|^2}{\sigma_0^2},  \forall n \in \mathcal{N}_{2h},
	\end{align} 
\end{subequations}
where $\rho_{n}$, $\psi_{n}$, and $\zeta_{n}$, are auxiliary variables.

Thus, by replacing constraints \eqref{P3-i}--\eqref{P3-k} with \eqref{lem1-formul}, and using $\Omega^{[i]}(\omega,z)$ for approximating the left side of \eqref{P3-b}, \eqref{P3-d}, and \eqref{P3-f}, the optimization problem \eqref{P3} transforms into a standard convex semidefinite programming (SDP). This can be solved using numerical solvers, such as the SDP tool in CVX \cite{CVX}.

\section{Numerical Results}\label{Simulat}
In our simulations, we consider a $3 \times 3$ $\rm m^2$  subnetwork with 20 sensors representing a production module in a factory. Sensors and APs are  uniformly distributed within the subnetwork. Other key simulation parameters are summarized in Table \ref{Table 1}, except when noted otherwise.
The wireless channels between devices and APs, and between sAPs and pAPs, are modeled with independent, frequency-flat Rayleigh fading. The path loss model follows the factory and open-plan building channel model from \cite{Khosravirad}. We use a shadow fading model with a standard deviation of 7 dB, as per \cite{3GPP}.
We test cooperative schemes with different numbers of sAPs, including `1h' and `1 of $K$'.
The FL algorithm is simulated using the Matlab Machine Learning Toolbox for handwritten digit recognition. For this task, each user trains an FNN with 50 neurons using the MNIST dataset\cite{MNIST}. The loss function is the cross-entropy loss.
\begin{table}
	\small
	\renewcommand{\arraystretch}{1.3}
	\caption{Federated learning parameters setup}
	\centering
	\label{Table 1}
	\resizebox{\columnwidth}{!}{
		\begin{tabular}{|c|p{60mm}|}
			\hline
			Parameter description  &  Value \\
			\hline \hline
			Number of sAPs, $K$ &  Variant based on the scheme  \\ 	\hline
			transmit data size $s$ & 10 kbits  \\ 		\hline
               Local sample data size $D_n$ & 200 samples  \\ 		\hline
			Maximum completion time, $T^{\mathrm{th}}$ & 6 s \\ 		\hline
   Uplink delay requirements $T'$ &4 ms \\ 		\hline
	    	Bandwidth & 100 MHz \\ 		\hline
			Carrier frequency &     10 GHz	 \\ 		\hline
			$P_{\max}$ &  23 dBm \\ 		\hline
			Power spectral density of the AWGN &  -174 dBm/Hz  \\ 				\hline
     The effective switched capacitance, $\kappa$ \cite{Energy1}&  $10^{-28}$  \\ 		\hline
     maximum computation capacity, $f^{\max}$&  1 GHz \\ 		\hline
The number of CPU cycles,  $C_n$ \cite{Energy1} & $[1, 2]\times 10^4$ (cycles/sample)  \\ 		\hline
		\end{tabular}}
	\end{table}
 
\begin{figure} 
	\centering
	\subfloat[\label{fig3}]{%
		\includegraphics[width=0.5\linewidth]{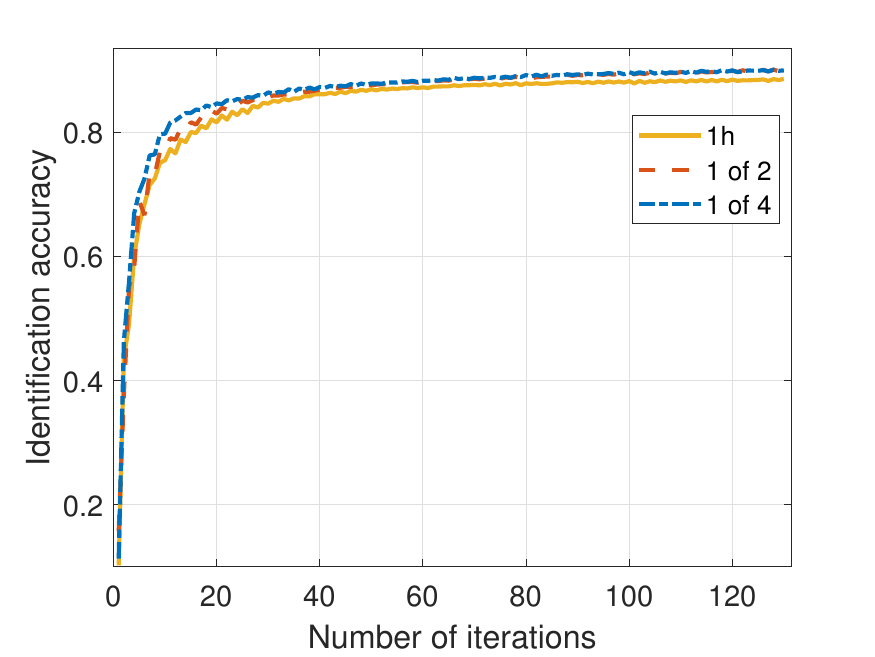}}
	\hfill
	\subfloat[\label{fig4}]{%
		\includegraphics[width=0.5\linewidth]{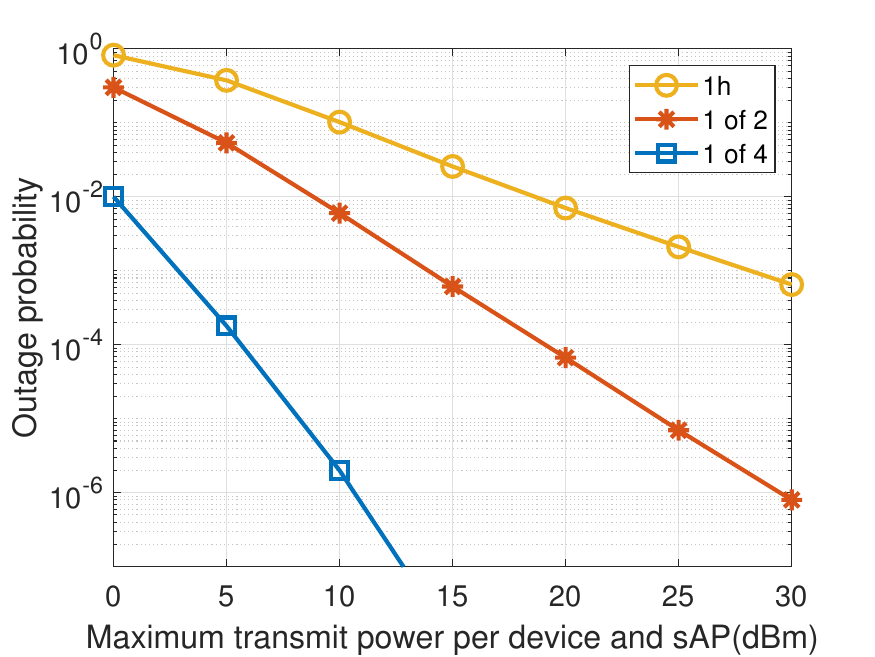}}
	\caption{(a) Identification accuracy against the number of iterations. (b) Outage probability versus the maximum transmit power per device and sAPs.}
\end{figure}

\begin{figure}
	\centering
	\includegraphics[width=.9\linewidth]{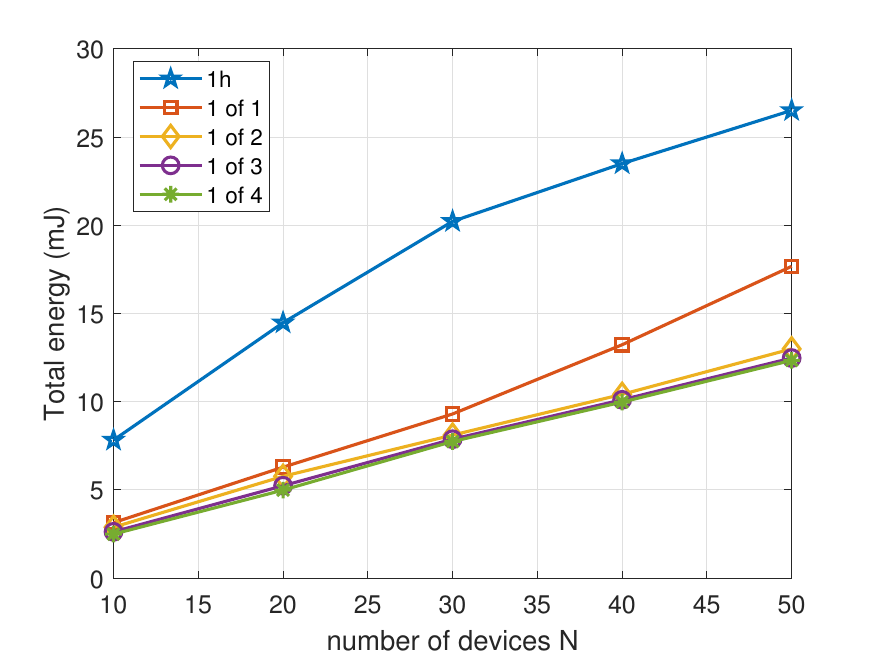}
	\caption{Total energy versus number of devices $N$ for different communication schemes with fixed target accuracy.}
	\label{fig5}	
\end{figure}

Fig. \ref{fig3} illustrates the identification accuracy as a function of the number of iterations for different transmission protocols. Because of the delay constraint defined in \eqref{P1}, single-hop transmission may not always meet this requirement and is therefore excluded from training in some scenarios. As a result, our proposed cooperative communication method demonstrates superior performance compared with single-hop transmission.
Fig. \ref{fig4} shows that the system outage performance improves as $P_{\max}$ increases. This is because higher transmit power at users and relays reduces latency during model upload. Additionally, having more sAPs reduces system outage probability by providing more options for selecting the strongest transmission link.
In Figure \ref{fig5}, we fixed the target accuracy ($I_0 = 130$, $I_n = 1$) and plotted total energy against the number of devices $N$ (ranging from 10 to 50), comparing different communication schemes. The results indicate that cooperative transmission outperforms single-hop transmission. Notably, the `1 of 4' scheme uses at least two times less energy than single-hop transmission.

\section{Conclusion}\label{conc}
In this paper, we considered an energy-efficient federated
learning framework where devices upload their locally
trained models in single-hop or two-hop transmission. We
proposed an efficient parameter optimization algorithm to
jointly optimize system parameters, such as the operating
frequency of each device, the  single-hop and two-hop devices and the associated relays, and the
transmit power. The proposed
SPCA algorithm can reasonably manage the energy
resources by balancing communication and local training
costs. 
Our experiments demonstrated that using four sAPs in federated learning can reduce total energy consumption by at least half.



\begin{thebibliography}{99}

\bibitem{Nguyen}
D. C. Nguyen \textit{et al.}, ``Federated learning for industrial internet of things in future industries,'' \textit{IEEE Wireless Commun.}, vol. 28, no. 6, pp. 192--199, Dec. 2021.

\bibitem{Hu}
Y. Hu \textit{et al.}, ``Industrial internet of things intelligence empowering smart manufacturing: A literature review,'' \textit{IEEE Internet Things J.}, vol. 11, no. 11, pp. 19143--19167, Jun. 2024.

\bibitem{Aouedi}
O. Aouedi \textit{et al.}, ``A survey on intelligent internet of things: Applications, security, privacy, and future directions,'' \textit{IEEE Commun. Surveys Tuts.}, vol. 27, no. 2, pp. 1238--1292, Apr. 2025.

\bibitem{Huang}
Y. Huang \textit{et al.}, ``Efficient and privacy-preserving authentication for federated learning in industrial internet of things data sharing application,'' \textit{IEEE Internet Things J.}, vol. 12, no. 9, pp. 11652--11663, May 2025.

\bibitem{Zhao}
Z. Zhao \textit{et al.}, ``Federated learning with non-IID data in wireless networks,'' \textit{IEEE Trans. Wireless Commun.}, vol. 21, no. 3, pp. 1927--1942, Mar. 2022.

\bibitem{Zhan}
Y. Zhan \textit{et al.}, ``A survey of incentive mechanism design for federated learning,'' \textit{IEEE Trans. Emerg. Topics Comput.}, vol. 10, no. 2, pp. 1035--1044, Mar. 2022.

\bibitem{Gilberto}
G. Berardinelli \textit{et al.}, ``Extreme communication in 6G: Vision and challenges for `in-X' subnetworks,'' \textit{IEEE Open J. Commun. Soc.}, vol. 2, pp. 2516--2535, 2021.

\bibitem{relay1}
X. Li, R. Fan \textit{et al.}, ``Energy-efficient resource allocation for mobile edge computing with multiple relays,'' \textit{IEEE Internet Things J.}, vol. 9, no. 13, pp. 10732--10750, Jul. 2022.

\bibitem{FL1}
B. McMahan \textit{et al.}, ``Communication-efficient learning of deep networks from decentralized data,'' in \textit{Proc. Int. Conf. Artif. Intell. Stat. (AISTATS)}, Apr. 2017, pp. 1273--1282.

\bibitem{Energy1}
Z. Yang \textit{et al.}, ``Energy-efficient federated learning over wireless communication networks,'' \textit{IEEE Trans. Wireless Commun.}, vol. 20, no. 3, pp. 1935--1949, Mar. 2021.

\bibitem{Hashempour}
H. R. Hashempour \textit{et al.}, ``Power-efficient cooperative communication within IIoT subnetworks: Relay or RIS?,'' \textit{IEEE Internet Things J.}, vol. 12, no. 9, pp. 12483--12500, May 2025.

\bibitem{CVX}
M. Grant and S. Boyd, ``CVX: MATLAB software for disciplined convex programming, version 2.1,'' [Online]. Available: http://cvxr.com/cvx, Mar. 2014.

\bibitem{Khosravirad}
S. R. Khosravirad, H. Viswanathan, and W. Yu, ``Exploiting diversity for ultra-reliable and low-latency wireless control,'' \textit{IEEE Trans. Wireless Commun.}, vol. 20, no. 1, pp. 316--331, Jan. 2021.

\bibitem{3GPP}
3GPP, ``Study on channel model for frequencies from 0.5 to 100 GHz,'' TR 38.901, v17.0.0, 2022.

\bibitem{MNIST}
Y. LeCun, ``The MNIST database of handwritten digits,'' [Online]. Available: http://yann.lecun.com/exdb/mnist/, Sep. 2020.

\end{thebibliography}
\end{document}